\documentstyle[multicol,epsf,aps]{revtex}

\begin{document}
             
\draft
\title{Anomalous Hall effect as a probe of the chiral order in spin glasses
}
\author{Hikaru Kawamura}
\address{Department of Earth and Space Science, Faculty of Science,
Osaka University, Toyonaka 560-0043,Japan}
\address{}
\date{\today}
\maketitle
\begin{abstract}
Anomalous Hall effect arising from the noncoplanar spin configuration
(chirality) is discussed as a probe of the chiral order in spin glasses.
It is shown that the  Hall coefficient yields direct
information about the
linear and nonlinear chiral susceptibilities of the spin sector,
which has been hard to obtain experimentally
from the standard magnetic measurements.
Based on the chirality scenario of spin-glass transition,
predictions are given on the behavior of the Hall resistivity of
canonical spin glasses.
\end{abstract}
\begin{multicols}{2}
\narrowtext

For decades,
spin glasses have been  extensively studied as a prototype of
``complex'' systems characterized by both `frustration' and `randomness'
\cite{reviewSG}.
Among a wide variety of spin-glass (SG) materials,
most familiar and well-studied is perhaps
the so-called canonical SG,
a dilute noble metal/3d transition metal alloys.
In canonical SG, the interaction between localized moments
is the RKKY interaction which is mediated by conduction electrons via
the $s-d$ exchange coupling $J_{sd}$. Oscillating nature of the RKKY
interaction with distance,
combined with spatially random arrangement of localized moments,
gives rise to frustration and randomness.
Since the RKKY interaction is isotropic in spin space,  canonical
SG like many other SGs is nearly isotropic in spin space,
and is expected to be well modeled by the Heisenberg model.
Weak magnetic anisotropy is mostly due to
the Dzyaloshinski-Moriya (DM) interaction caused
by the combined effect
of the $s-d$ coupling and the spin-orbit interaction.
Nearly isotropic character of the magnetic interaction in
canonical SG is in apparent contrast to
most of theoretical approaches which have been based on the Ising model
describing the extremely anisotropic limit\cite{reviewSG}.

Experimentally, it is  now well established that typical SG magnets
including canonical SG exhibit an equilibrium phase transition
at a finite temperature and there exists a thermodynamic SG
phase. True nature of the SG transition and of
the SG ordered state,
however, still remains elusive in spite of extensive studies\cite{reviewSG}.

Although standard theories of the SG order
invoke the Ising model as a minimal model,
a scenario very different from the standard picture was proposed by the
present author, which may be called
a chirality scenario\cite{Kawamura92,Kawamura98}.
%in order to solve some of the puzzles
%concerning the experimentally observed SG transition
In this scenario, {\it chirality\/}, which is a multispin quantity
representing  the sense or the handedness of local
noncoplanar structure of Heisenberg spins,
plays an essential role. The local chirality may be defined
for three neighboring Heisenberg spins by the scalar,
\begin{equation}
\chi_{ijk}=\vec S_i\cdot \vec S_j\times \vec S_k.
\end{equation}
The chirality defined above is often called a scalar chirality:
%, to distiguish it from the vector chirality, the latter being
%defined for two neighboring spins by
%$\kappa \propto \vec S_i\times \vec S_j$.
It takes a nonzero value when the three spins
span the noncoplanar configuration in spin space, whose sign representing the
handedness of such noncoplanar spin configuration. 
%In particular,
%it vanishes for collinear or planar spin configuration.

The chirality scenario of  SG transition consists of the two parts
\cite{Kawamura92,Kawamura98}:
In a fully isotropic Heisenberg SG,
it claims the occurrence of a novel {\it chiral-glass\/}
ordered state in which only the chirality exhibits a glassy long-range
order keeping the Heisenberg spin paramagnetic (spin-chirality
decoupling). In a weakly anisotropic Heisenberg SG, the
scenario claims that the
weak random magnetic anisotropy
``recouples'' the spin to the chirality,
and the chiral-glass order of the fully isotropic system
shows up in the spin sector as the
standard SG order via the magnetic anisotropy (spin-chirality recoupling).
In other words, experimental SG transition and SG ordered state
are the chiral-glass transition and
chiral-glass ordered state of the fully isotropic system
``revealed'' by the weak random magnetic anisotropy inherent to real SG.

%Note that this point is very different from the
%conventional views  where it is implicitly assumed that
%all essential features of experimental SG transition can be captured
%by understanding the properties of the Ising EA model.

So far,  experimental test of the chirality scenario remains
indirect. This is mainly due to the experimental difficulty in
directly measuring the  chirality. %due to its multispin (cubic) nature.
A possible clue to overcome this difficulty was recently
found via the study of electron transport properties of certain magnets
in which conduction electrons interact with the core spins possessing
chiral degrees of freedom. In particular,
it has been realized that under appropriate conditions
a chirality contribution shows up in the anomalous part
of the Hall effect.
This was first pointed out in the strong coupling case
where the conduction electrons are
strongly coupled with core spins via the Hund coupling, as in the
cases of manganites\cite{Ye,Chung} and in frustrated kagom\'e
\cite{Nagaosa}, or pyrochlore
ferromagnet\cite{Taguchi1,Taguchi2} (see also Ref.\cite{Sato}). In the weak
coupling case
which is more relevant to canonical SG,
chirality contribution to the anomalous Hall effect
was examined by Tatara and Kawamura\cite{TK}. By applying the
linear response theory and the perturbation expansion to
the standard $s-d$ Hamiltonian, these authors derived the
chirality contribution to the Hall resistivity.
%, and pointed out
%its potential importance in detecting the chiral order in SG.

Under such circumstances,
the purpose of the present letter is twofold.
First, on the basis of the
formula derived in Ref.\cite{TK}, I wish
to explore in some detail the properties of
the Hall resistivity at the SG transition with particular interest in
the chirality contribution, and propose the way
to extract information about the chiral order of the spin sector
from the experimental data.
Second, on the basis of
the aforementioned chirality theory of SG transition,
I present several predictions on
the expected behavior of the anomalous part of
the Hall resistivity, which might serve as an experimental test of
the chirality theory. In the following, I will discuss these two issues
successively.

Let us
begin with summarizing the perturbative results for the Hall 
resistivity\cite{TK}, 
which will form the basis
of the following analysis. 
Conduction electrons on the lattice with $N$ sites are coupled
with core spins (assumed to be classical and fixed) 
via the standard $s-d$ exchange interaction $J_{sd}$, and are also 
scattered by normal impurities.
%
%\begin{equation}
%{\cal H}=\sum _{\vec k,\sigma'=\pm}\epsilon _{\vec k \sigma'}c^\dagger
%_{\vec k \sigma'}c_{\vec k \sigma'} + \frac{J_{sd}}{N}\sum_{\vec k\vec k'}\vec
%S_{\vec k'-\vec k}(c_{\vec k'}\vec\sigma c_{\vec k}) + {\cal H}_{{\rm imp}},
%\end{equation}
%
%where $\epsilon _{\vec k \sigma'}$ is the single-particle energy of
%conduction electron, $\sigma ^\alpha $ ($\alpha =x,y,z$) are Pauli matrices,
%and $N$ is the total number of lattice sites. 
%
%
%The core spins $\vec S_i$ are assumed to be classical and fixed.
%Scattering by normal impurities is represented by ${\cal H}_{{\rm imp}}$:
%See ref.\cite{TK} for further details.
%
Assuming the weak-coupling regime in which $J_{sd}$ is
smaller than the Fermi energy $\epsilon _F$, first nonzero contribution to the
Hall conductivity comes from the third-order term in the perturbation,
which can be recast into the Hall resistivity as
\begin{equation}
\rho_{xy}^{({\rm chiral})} = 54\pi^2\chi_0
\left(\frac{J_{sd}}{\epsilon_F}\right)^2J_{sd}\tau \rho_0 
= CJ_{sd}^3\chi_0, \\
\end{equation}
\begin{eqnarray}
\chi_0 = \frac{1}{6Nk_F^2}\sum _{ijk}\chi_{ijk}
\left[ \frac{(\vec r_{ij}\times \vec r_{jk})_z}{r_{ij}r_{jk}}
I'(r_{ij})I'(r_{jk})I(r_{ki})\right. \nonumber \\
\left. +({\rm two\ permutations}) \right], 
\end{eqnarray}
where $\rho_0$ is the Boltzmann resistivity, $\tau$ the mean collision time,
$k_F$ the Fermi wavenumber,
and $\chi_{ijk}$ represents the local chirality defined by eq.(1),
$\vec r_{ij}=\vec r_i-\vec r_j$,
{\it etc\/}, with $r_{ij}\equiv |\vec r_{ij}|$, {\it etc\/}.
$I(r)$ represents a function decaying as $I(r)= \frac{\sin k_Fr}{k_Fr}
e^{-r/2\ell}$, with $\ell$ being the electron mean free path,
and $I'(r)=\frac{{\rm d}I(r)}{{\rm dr}}$.
One sees from eq.(3) that $\chi _0$ is a total (net) chirality,
while the factor in the square bracket in eq.(3) specifies the coupling between
the spin space and the real space. In canonical SG,
$J_{sd}$ is positive. The coefficient $C$ is
%calculated to be $C=\frac{1}{e^2}\frac{2V}{N}(\frac{m}{k_F})^2$ and is 
positive in the single-band approximation, but more generally,
its sign would depend on the detailed band structure of the material.
%Anyway, in the weak coupling regime relevant to canonical SG,
%the leading contribution to the Hall resistivity is proportional
%to the toal chirality.

By contrast, 
conventional theories of the anomalous Hall effect have attributed its origin
to the spin-orbit interaction $\lambda $ and a finite magnetization $M$
\cite{KL,Smit,Luttinger}, {\it i.e.\/},
mechanisms known as the skew scattering or the side jump.
Note that the chirality contribution is independent of these
conventional ones. Taking account of the
conventional terms within the perturbation scheme, the anomalous part of the
Hall resistivity has been given by
\begin{eqnarray}
\rho_{xy} &=& -\lambda M(A\rho+B\rho^2)+CJ_{sd}^3\chi_0,\nonumber \\
          &=& -M(\tilde A\rho+\tilde B\rho^2)+\tilde C\chi_0,
\end{eqnarray}
where $\rho$ is the longitudinal resistivity $\rho=\rho_{xx}$,
$A$ and $B$ are constants both positive within the single-band
approximation\cite{TK}, and $\tilde A=A\lambda$, $\tilde B=B\lambda$,
$\tilde C=CJ^3$.

Since Heisenberg spins
are frozen in a spatially random manner  in the SG ordered state,
the sign of the local chirality appears
randomly, which inevitably leads to the vanishing total chirality
in the bulk, $\chi_0=0$. %, due to  obvious cancellation effect.
It thus appears that the chirality-driven
anomalous Hall effect vanishes in bulk SG samples.
% even though a finite contribution exists locally.  
In the strong coupling case, however, 
a mechanism out of this cancellation
was proposed by Ye {\it et al\/}\cite{Ye}. These authors
pointed out that the spin-orbit interaction $\lambda$ in the presence of a
net magnetization $M$ contains a term of the form
\begin{equation}
{\cal H}_{so}\approx \tilde DM\chi_0,
\end{equation}
which, in the spin Hamiltonian, couples the total chirality to the total
magnetization. In the weak coupling regime relevant to canonical SG,
a term of the form (5) with $\tilde D=D\lambda (J_{sd}/\epsilon_F)^2(J\tau)^2$ 
was also derived
perturbatively by taking the
electron trace of the spin-orbit interaction\cite{TK}.
The sign of the coefficient
$D$ generally depends on the detailed band structure\cite{TK},
while Ye {\it et al\/} argued
that $\tilde D$ should be positive\cite{Ye}. In any case, 
a crucial observation here
is that the weak chiral symmetry-breaking term (5)
guarantees a net
total chirality to be induced if the sample is magnetized. Net
magnetization may be generated spontaneously (ferromagnet or reentrant SG)
or induced by applying external fields. 

I now go on to analyze the behavior of the anomalous
Hall resistivity of SG based on eqs.(4) and (5).
I assume for the time being
that the system does not possess a spontaneous magnetization, namely,
the magnetization is the one induced by external magnetic field $H$.

The quantities playing a crucial role
in the following analysis are the linear and
nonlinear {\it chiral\/} susceptibilities, defined by,
\begin{equation}
X_\chi=\left.\frac{{\rm d}\chi_0}{{\rm d}H_\chi}\right |_{H_\chi=0},\ \
X_\chi^{nl}=\frac{1}{6} 
\left.\frac{{\rm d}^3\chi_0}{{\rm d}H_\chi^3}\right |_{H_\chi=0},
\end{equation}
where $H_\chi$ is the ``chiral field'' conjugate to the net chirality
$\chi_0$, {\it i.e.\/}, $H_\chi$ couples to the net chirality
as $-H_\chi \chi_0$ in the spin Hamiltonian. Note that the chiral
symmetry-breaking
interaction discussed above, eq.(5), has exactly this form with
$H_\chi=-\tilde DM$.

With use of the linear and nonlinear chiral susceptibilities,
the total chirality can be written as
\begin{equation}
\chi_0=-X_\chi (\tilde DM)-X_\chi^{nl}(\tilde DM)^3+\cdots .
\end{equation}
If one substitutes this into eq.(4), one gets the chirality contribution
to the anomalous part of the Hall resistivity as
\begin{equation}
\rho_{xy}^{({\rm chiral})}=-\tilde C\tilde DM
[X_\chi+X_\chi^{nl}(\tilde DM)^2+\cdots].
%=CD[X_M+X_M^{nl}H^2+\cdots][X_\chi+X_\chi^{nl}(\tilde DM)^2+\cdots].
\end{equation}
Including the contributions of
the skew scattering and the side jump,
the Hall coefficient $R_s$, defined as the anomalous Hall resistivity divided
by the magnetization $R_s=\rho_{xy}/M$, is  given by
\begin{equation}
R_s= -\tilde A\rho-\tilde B\rho^2-\tilde C\tilde D
[X_\chi+X_\chi^{nl}(\tilde DM)^2+\cdots ].
\end{equation}
The total Hall resistivity contains in addition the contribution from
the normal part, which, we assume throughout this analysis, has properly
been subtracted.
One can immediately see from eq.(9) that the anomalous 
Hall coefficient $R_s$ carries
information of the chiral susceptibilities. In particular, in
the linear regime where
the magnetization is sufficiently small and the Hall resistivity
is proportional to $M$, the chirality contribution to
$R_s$ is proportional to the linear chiral susceptibility $X_\chi$.

In the standard measurements of Hall resistivity,
dc magnetic field is applied,
either in field-cooling (FC) or zero-field-cooling (ZFC) conditions.
%The dc magnetization is generally given in terms of
%the linear and nonlinear magnetic susceptibilities
%(to be distinguished from the chiral susceptibilities)
%$X_m=\frac{{\rm d}M}{{\rm d}H}$ and
%$X_m^{nl}=\frac{{\rm d}^3M}{{\rm d}H^3}$ as $M=X_mM+X_m^{nl}M^3+\cdots $.
As is well-known, at the SG transition
temperature $T=T_g$, the linear magnetic susceptibility
$X_m=\left.
\frac{{\rm d}M}{{\rm d}H}\right | _{H=0}$ (to be distinguished from the
linear chiral susceptibility)
exhibits a cusp
accompanied by the
onset of deviation between
the FC and ZFC susceptibilities\cite{reviewSG}.
Sharp cusp of the linear susceptibility at $T=T_g$
is known to be rounded off by
applying weak external magnetic fields, which is also
manifested in the well-known negative divergence
of the nonlinear magnetic susceptibility
$X_m^{nl}=\frac{1}{6}
\left.\frac{{\rm d}^3M}{{\rm d}H^3}\right | _{H=0}$ at $T=T_g$\cite{reviewSG}.
By contrast, the resistivity $\rho$ of canonical SG exhibits no detectable
anomaly at $T=T_g$\cite{reviewSG}.

The Hall resistivity is generally given by the combination
of both the magnetic and chiral susceptibilities, as seen
from eq.(8) with $M=X_mH+X_m^{nl}H^3+\cdots $.
By contrast, one can extract
information solely about the chiral susceptibilities from
the Hall coefficient $R_s$ which is obtained by dividing the
Hall resistivity by the magnetization measured simultaneously
or in the same condition. Here, note that the
magnetization of SG exhibits a singular behavior at $T=T_g$.
Furthermore, by examining the $M$-dependence of the Hall
coefficient in the nonlinear regime, one can  extract
information about the nonlinear chiral susceptibility.
%In practice, the separation procedure of the normal part might
%cause some ambiguity.
Anyway, in contrast to the standard
magnetic susceptibilities measurable by the conventional technique,
information about the chiral susceptibilites
have so far been hard to get experimentally,  and if measurable as above,
would be very valuable.

Next, I with to give predictions on the behavior of
the anomalous Hall coefficient
based on the chirality scenario of SG transition
\cite{Kawamura92,Kawamura98}.
Chirality scenario predicts that, in both isotropic and weakly
anisotropic Heisenberg SGs, the chirality behaves as an order parameter of
the transition (chiral-glass transition).  
The singular part of the free
energy should satisfy the scaling form,
\begin{equation}
f_s\approx |t|^{2\beta_\chi+\gamma_\chi}F_\pm\left (\frac{H_\chi^2}
{|t|^{\beta_\chi+\gamma_\chi}}\right ),
\end{equation}
where $\beta_\chi$ and $\gamma_\chi$ are the chiral-glass order parameter
and chiral-glass susceptibility exponents, respectively,
$t\equiv (T-T_g)/T_g$ is a reduced temperature, and $F_\pm(x)$ is a scaling
function either above ($+$) and below ($-$) $T_g$.  Numerical
estimates give $\beta_\chi\sim 1$ and $\gamma _\chi \sim 2$
\cite{HukuKawa,ImaKawa}.
%In the following, I put .
By differentiating eq.(10) with respect to $H_\chi$ and putting $H_\chi=0$,
one sees
that at $T=T_g$ the linear chiral susceptibility exhibits a cusp-like
singularity while the nonlinear chiral susceptibility
exhibits a negative divergence,
\begin{equation}
X_\chi\approx c_0^{(\pm)}|t|^{\beta_\chi}+b_0(t),\ \  X_\chi^{nl}\approx
c_2^{(\pm)}|t|^{-\gamma_\chi}+b_2(t),
\end{equation}
with $\beta_\chi \sim 1$ and $\gamma_\chi \sim 2$,
where $c_0^{(\pm)}<0$ and $c_2^{(\pm)}<0$ are constants describing either above
or below $T_g$, while $b_0(t)>0$ and $b_2(t)$
represent regular terms coming from the nonsingular part.
%The negative divergence of $X_\chi^{nl}$
%can be derived on general groudns, {\it e.g.\/},
%by applying the Ginzburg-Landau-type analysis of Ref.\cite{Suzuki}.
Concerning the standard magnetic susceptibilities,
the chirality scenario predicts that, in the realistic case of weakly
anisotropic Heisenberg SG, $X_m$ and $X_m^{nl}$ exhibit the same
singularities as  $X_\chi$ and $X_\chi^{nl}$, which are
caused by the spin-chirality
recoupling  due to the random magnetic anisotropy.
(In the hypothetical
limit of zero anisotropy, because of the spin-chirality decoupling in the
isotropic system, $X_m$ and $X_m^{nl}$ are predicted to exhibit less
singular behaviors very different from those of
$X_\chi$ and $X_\chi^{nl}$ .
But, after all, certain amount of
anisotropy is inevitable in real SG, which eventually causes the
spin-chirality recoupling.)

The anomalous Hall coefficient of SG should be dominated by the
singular behaviors of the chiral susceptibilities, since the
first and second terms of the r.h.s. of eq.(9) can be regarded as a regular
background because of the nonsingular behavior of $\rho$.
By combining the observations above, the
following predictions follow.
(i) The linear part of $R_s$,
which is $R_s$ itself in the linear regime where
$R_s$ is proportional to the magnetization $M$,
exhibits a cusp-like anomaly at $T=T_g$, possibly accompanied by the
onset of the deviation between the FC and ZFC results.
This cusp-like singularity
is rounded off in the presence of a finite magnetization. (ii) The nonlinear
part of $R_s$, which can be extracted by examining the $M$-dependence
of $R_s$ in the nonlinear regime,
exhibits a divergence at $T=T_g$ characterized by the
exponent $\gamma_\chi \sim 2$ which is equal to the standard
nonlinear susceptibility exponent $\gamma$. (iii) The chiral part
of $R_s$, obtained by properly subtracting the background due to
the possible contribution of the skew scattering and the side jump
{\it etc\/},
is expected to obey the scaling form,
\begin{equation}
R_s^{({\rm chiral})}\approx |t|^{\beta_\chi}
G_\pm(\frac{M^2}{|t|^{\beta_\chi+\gamma_\chi}}),
\end{equation}
where $G_\pm(x)$ is a scaling function either above or below
$T_g$. The subtraction of the background might be performed
by analyzing the temperature dependence of $R_s$ based on eq.(9), 
using the data
of the resistivity $\rho$. 
(iv) The sign of the Hall resistivity depends on the signs and the
relative magnitudes
of constants $\tilde A$, $\tilde B$, $\tilde C$ and $\tilde D$ ($X_\chi$
is positive by definition). Hence, the sign of $\rho_{xy}$
seems nonuniversal, depending
on the band structure of the material. In cases where the single-band
approximation and the naive argument of Ref.\cite{Ye} concerning the sign of
$\tilde D$ are valid, one has
$\tilde C>0$ (with $J_{sd}>0$)
and $\tilde D>0$, which means $\rho_{xy}^{({\rm chiral})}$
is negative in canonical SG. This seems consistent with experiment\cite{McAlister,Barnard}. 
The cusp-like singularity
was observed there 
at least in the Hall resistivity\cite{McAlister,Barnard}, 
consistent
with the present result.
%Thus, measurements of the Hall
%resistivity, resistivity and magnetization in
%applied magnetic fields would
%enable one to extract information about the chiral order in canonical SG,
%and to test the chirality theory of  SG transition.

Note that even the conventional mechanism of the anomalous Hall effect
(the skew-scattering or the side-jump mechanism) predicts that
the Hall resistivity exhibits
a cusp-like singularity at $T=T_g$, which is a reflection of the
cusp-like singularity of the magnetic susceptibility.
However, the conventional mechanism also predicts that
the Hall coefficient $R_s$ behaves in a nonsingular manner at  $T=T_g$ 
as $\rho$ and $\rho^2$.
A highly nontrivial issue 
is then whether the Hall coefficient, not just the
Hall resistivity, exhibits an anomaly at $T=T_g$. If
singular behavior is observed in $R_s$, it is likely to be of
chirality origin.
%As mentioned,
%the chirality scenario of  SG transition predicts a cusp-like behavior
%for the linear chiral susceptibility,
%and most remarkably, a divergent behavior for the
%nonlinear chiral susceptibility.

One can give a rough order estimate of $\rho_{xy}^{({\rm chiral})}$.
In typical canonical SG like AuFe and CuMn, $J_{sd}/\epsilon_F$ and $J\tau$
are of order $10^{-1}$ and $10^0$, respectively. Then, from 
eq.(2),
$\rho_{xy}^{({\rm chiral})}$
is estimated to be of order $M\chi_0$  in units of $\rho_0$.
Since the magnitude of the 
chiral symmetry-breaking interaction (5) is of order the 
DM interaction, the induced chirality $\chi_0$
is of order [DM interaction]/[RKKY interaction].
This ratio is a material dependent parameter, being small for CuMn, say, 
$10^{-2}$, and relatively large for AuFe, say, $10^{-1}$ or more.
Thus, if the sample is magnetized 10\% 
of the saturation value, one expects for AuFe $\rho_{xy}^{({\rm chiral})}$ 
of order percents of $\rho_0$ 
or even more. For more isotropic materials like CuMn and AgMn,
the chiral contribution would be reduced being proportional to the strength
of the DM interaction.

Finally, I wish to discuss the reentrant SG with a spontaneous magnetization.
With decreasing temperature, the reentrant SG exhibits
successive transitions,
first from para to ferro at $T=T_c$, then from ferro to reentrant SG at
$T=T_g$. In the ferromagnetic phase at $T_c>T>T_g$, the system
exhibits a spontaneous magnetization without the chiral order,
whereas at $T=T_g$ the glassy chiral order sets in accompanied with
the random spin canting in the transverse direction, which coexists
with the spontaneous magnetization in the longitudinal direction.
The present result for the Hall resistivity also
applies to such reentrant SG around $T=T_g$,
only if $M$ is treated as including the spontaneous
magnetization.
In the ferromagnetic state,
the chiral susceptibility should remain
small due to the absence of chiral order, and
the anomalous Hall resistivity should be dominated by
the conventional contributions from the skew scattering and/or side-jump.
Below $T=T_g$, additional contribution from the
chiral order sets in, giving rise to a cusp in the
Hall resistivity at $T=T_g$. As often observed under FC conditions,
magnetization of reentrant SG
is saturated at temperatures far above $T_g$, with very little
anomaly at $T=T_g$.  In such a case, if anomaly is observed
in the FC mode
in the Hall resistivity at $T=T_g$, this
can be identified as arising from the chirality.
Indeed, a cusp-like  anomaly at $T=T_g$ was
recently observed in  manganite reentrant SG
La$_{1.2}$Sr$_{1.8}$Mn$_2$O$_7$ \cite{Chung2} and in reentrant SG alloy
Fe$_{1-x}$Al$_x$\cite{Sato2}, 
suggesting that the
observed anomaly is of  chirality origin.

In summary, I examined the Hall resistivity of
canonical SG, and have found that the Hall coefficient gives
information about the linear and nonlinear chiral susceptibilities
of SG. Based on the chirality scenario,
predictions were given on the behavior of the Hall
coefficient of canonical SG.
% $R_s$ of canonical SG, in particular, the occurrence of
%a cusp in the linear part of $R_s$ as well as the occurrence of a
%divergence in the nonlinear part of $R_s$ with an exponent $\gamma _\chi
%\sim 2$. 
I hope the present work
will stimulate further experimental activities on the chiral
order and the Hall resistivity of SG and related materials.

The author is thankful to Dr. M. Sato, Dr. S. Kawarazaki, 
Dr. T. Taniguchi, and Dr. G. Tatara for useful discussion.
This work was supported by Priority Areas Grants from the Ministry of
Education, Culture, Sports, Science and Technology, Japan.

\end{multicols}

\end{document}